\documentclass[%
 reprint,
 amsmath,amssymb,
 aps,
]{revtex4-2}

\usepackage{graphicx}% Include figure files
\usepackage{dcolumn}% Align table columns on decimal point
\usepackage{bm}
\usepackage{xcolor}
\begin{document}
\preprint{APS/123-QED}
\title{Green's functions in quantum mechanics courses}

\author{William J. Herrera$^{1}$}
\author{Herbert Vinck-Posada$^{1}$}
\author{Shirley G\'omez P\'aez$^{1,2}$}
\affiliation{$^{1}$Departamento de F\'isica, Universidad Nacional de Colombia, 111321, Bogot\'a, Colombia}
\affiliation{$^{2}$Departamento de F\'{\i}sica, Universidad el Bosque, Bogot\'a, Colombia.}
\begin{abstract}
Green’s functions in Physics have proven to be a valuable tool for understanding fundamental concepts in different branches, such as electrodynamics, solid-state and many-body problems. In quantum mechanics advanced courses, Green's functions usually are explained in the context of the scattering problem by a central force. However, their use for more basic problems is not often implemented. The present work introduces Green's Function in quantum mechanics courses with some examples that can be solved with essential tools. For this, the general aspects of the theory are shown, emphasizing the solution of different fundamental issues of quantum mechanics from this approach. In particular, we introduce the time-independent Green's functions   and the Dyson equation to solve problems with an external potential. As examples, we show the scattering by a Dirac delta barrier, where the reflection and transmission coefficients are found. In addition, the infinite square potential well energy levels, and the local density of states, are calculated.
\end{abstract}
%
%\pacs{42.50.Pq, 71.36.+c, 73.43.Nq}
%
\maketitle
\section{\label{sec:level0}Introduction}
Green's function method to solve problems in different areas of physics is done at the undergraduate or graduate level. For example, the usual thing in physics programs is introducing Green's functions to solve inhomogeneous differential equations, such as the Poisson's equation, inhomogeneous wave equation, inhomogeneous heat equation \cite{Jackson:1972, Schwartz:1972, Griffiths:2017,Asmar:2016, RubinLan:2004, Baym:1993, Hameka:2004}. On the other hand, in advanced topics, such as a many-body problem or solid-state physics, Green's functions are introduced to solve more complex problems that usually require concepts of the second quantization \cite{Rickayzen:1980, Economou:2006, Bruss:2004, Fetter-Walecka:2003, Kittel:1969, Zagoskin:1998, March:1995, Mahan:1990, Raimes:1972}. Moreover, in quantum mechanics courses, the exposition is usually posed in the context of scattering by a central potential, which implies making theoretical developments dependent and independent of time  and the use of spherical or cylindrical coordinates \cite{Griffiths:2018, Sakurai:2017, Cohen:1991, Shift:1968}. These facts usually lead to Green's functions in quantum mechanics not being usually exposed in undergraduate courses, which does not allow undergraduate and postgraduate students to be aware of Green's functions in the context of quantum mechanics.  However, sometimes it involves realizing more elaborate calculations or using the complex variable \cite{Cohen:1991v2,Tsaur:1972, Lucas:1968, Lawson:1972, Whitten:1975, Byrd:1976, Prato:1983, Sukumar:1990, Schmalz:2010, Kamal:1984, Anderson:1989}, and other works show methods to find the Green's function in specific problems of quantum mechanics \cite{Sukumar:1990, Lessie:1986, Shao:2016}. In this work, we make an approach that can help introduce the concept and service of  Green's function in intermediate or advanced quantum mechanics courses. First, we present the formalism of Green's functions and how we can use it for the time-independent Schr\"{o}dinger equation. Later, we explain Green's function of  a free particle and derive the Dyson equation when the system is perturbed with a scalar potential. In particular, we consider a Dirac delta potential, where we find the Green's function for both reflection and transmission coefficients. Likewise, it illustrates how to find the Green's function of an infinite square potential well, and from it, we can calculate the spectrum of energy and the local density of states (LDOS).

\section{\label{sec:level2} Green's functions for the Schr\"{o}dinger and Dyson equations.}

Before starting with the implementation of the Green's function for the Schr\"{o}dinger equation, let us do a brief review of the Green's function associated with a linear operator $\hat{L}_{x}$ in the $x$ coordinate representation, 

\begin{equation}
\left( \lambda -\hat{L}_{x}\right) \phi \left( x\right) =f\left( x\right).
\label{MFG_eq1}
\end{equation}
We wish to find the inverse operator $\left( \lambda -\hat{L}\right) ^{-1}$, such that 
$\phi =\left( \lambda -\hat{L}\right) ^{-1}f$. The nucleus of an integral $G(x,x^{\prime
}) $ will represent this inverse operator,
defined as
\begin{equation}
\left( \lambda -\hat{L}\right) _{x,x^{\prime }}^{-1}=G(x,x^{\prime }),
\end{equation}%
such that the integral solution of the equation (\ref{MFG_eq1}) can write, as
\begin{equation}
\phi (x)=\int G(x,x^{\prime })f(x^{\prime })dx^{\prime },
\label{MFG_sol integral}
\end{equation}
where $ G(x,x^{\prime }) $ is known as Green's function (GF). Now let us derive the differential equation that satisfies $ G(x,x^{\prime }) $, applying the operator $ \lambda-\hat{L} $ to the equation (\ref{MFG_sol integral}), in this manner
\begin{eqnarray}
\left( \lambda -\hat{L}_{x}\right) \phi (x) &=&\int \left( \lambda -\hat{L}%
_{x}\right) G(x,x^{\prime })f(x^{\prime })dx^{\prime } \\
&=&f(x).
\end{eqnarray}%
Therefore $G(x, x^{\prime}) $ must satisfy \qquad
\begin{equation}
\left( \lambda -\hat{L}_{x}\right) G(x,x^{\prime })=\delta (x-x^{\prime }).
\label{MFG_eq_dif_FG}
\end{equation}
Furthermore, Green's function coordinate representation satisfies an inhomogeneous differential equation with $f(x)= \delta (x - x^{\prime})$. From the solution of Eq. (\ref{MFG_eq_dif_FG}) we can  solve the equation (\ref{MFG_sol integral}). Using the
inverse operator notation gives that
\begin{equation}
...\left( \lambda -\hat{L}\right) ^{-1}=\int ...G(x,x^{\prime })dx^{\prime }.
\label{Eq_op_inv_inte}
\end{equation}
In such a way,
\begin{equation}
\left( \lambda -\hat{L}_{x}\right) \left( \lambda -\hat{L}_{x}\right)
^{-1}=\int \delta (x-x^{\prime })dx=1.
\end{equation}

Now, we consider the  time-independent Schr\"{o}dinger equation, 
\begin{equation}
H_{0}\left( \mathbf{r}\right) \psi (\mathbf{r})=E\psi _{0}(\mathbf{r}),
\end{equation}%
with
\begin{equation}
H_{0}\left( \mathbf{r}\right) =-\dfrac{\hbar ^{2}}{2m}\nabla ^{2}+V_{0}(%
\mathbf{r}).
\end{equation}%
Here, the subscript $ 0 $ refers to the potential $ V_ {0} $. Even if
this equation is homogeneous; we can define the associated Green's function, as
\begin{equation}
\left( E-H_{0}\left( \mathbf{r}\right) \right) g(\mathbf{r,r}^{\prime
})=\delta (\mathbf{r-r}^{\prime }),
\label{Eq_g_H_0}
\end{equation}%
whose solution we can write as,
\begin{equation}
g(\mathbf{r,r}^{\prime })=\left( E-H_{0}\left( \mathbf{r}\right) \right)
^{-1}\delta (\mathbf{r-r}^{\prime }).
\end{equation}
The function $ g (\mathbf{r, r}^{\prime}) $ depends on $ E $, which does not   explicitly notice. To the value of $ E $, we usually add or subtract a small imaginary part, $ E \rightarrow E \pm i \eta $, thus $ g(\mathbf{r, r}^{\prime}) $ has no poles on the real axis when $ E $ matches a set energy value of the system. When we add $%
i \eta $, the Green's function is called retarded ($r$) and for $ -i \eta $
advanced ($a$), see note \footnote{When it is calculated the Green's function depending on $t-t^{\prime}$ as the Fourier transform of $G^{a/r}(E)$ it is found that $G^{a/r}(t-t^{\prime})\propto \exp(-iE(t-t^{\prime)}/\hbar\pm\eta (t-t^{\prime)}))$), so in order to have convergence for the advanced  is necessary that $t^{\prime}>t$ while that for the retarded $t>t^{\prime}$}. In the following, only when explicitly needed we will refer to the advanced or retarded Green's function.

The Green's function $ g(\mathbf {r, r}^{\prime}) $ becomes significant when to the Hamiltonian $ H_{0} \left (\mathbf{r}\right) $ we add a potential $ V \left(\mathbf{r}\right) $, which involves solving
\begin{equation}
\left( H_{0}\left( \mathbf{r}\right) +V\left( \mathbf{r}\right) \right) \psi
(\mathbf{r})=E\psi (\mathbf{r}),
\end{equation}%
so, we can express it as
\begin{equation}
\left( E-H_{0}\left( \mathbf{r}\right) \right) \psi (\mathbf{r})=V\left( 
\mathbf{r}\right) \psi (\mathbf{r}).
\end{equation}

We can see this equation as an  ``inhomogeneous" Schr\"{o}dinger equation, where the external source is given by $ f \left(\mathbf{r}\right) = V \left(\mathbf{r}\right)\psi(\mathbf{r})$, we start from the fact that we know $ g(\mathbf{r, r}^{\prime}) $, and then  we can write  the solution for $ \psi (\mathbf {r}) $  as
\begin{equation}
\psi (\mathbf{r})=\psi _{0}\left( \mathbf{r}\right) +\int d\mathbf{r}%
^{\prime }g\left( \mathbf{r},\mathbf{r}^{\prime }\right) V\left( \mathbf{r}%
^{\prime }\right) \psi (\mathbf{r}^{\prime }).
\end{equation}
This is an integral equation to find $ \psi (\mathbf{r}) $ from $ g\left (\mathbf{r}, \mathbf{r}^{\prime}\right) $ however, it is possible to express $ \psi (\mathbf{r}) $ more directly from the Green's function of the perturbed system, to define $ G (\mathbf{r, r}^{\prime}) $, as
\begin{equation}
\left( E-H\right) G(\mathbf{r,r}^{\prime })=\delta (\mathbf{r-r}^{\prime }),
\label{Eq_G_dif}
\end{equation}
with $H=H_{0}\left( \mathbf{r}\right) +V\left( \mathbf{r}\right)$. 
Using Eq.(\ref{Eq_g_H_0}) to write $\delta (\mathbf{r-r}^{\prime })$, $G(\mathbf{r,r}^{\prime })$ can be written as
\begin{eqnarray}
G(\mathbf{r,r}^{\prime }) &=&\left( E-H\left( \mathbf{r}\right) \right)
^{-1}\delta (\mathbf{r-r}^{\prime })  \nonumber \\
&=&\left( E-H\left( \mathbf{r}\right) \right) ^{-1}\left( E-H_{0}\left( 
\mathbf{r}\right) \right) g(\mathbf{r,r}^{\prime })  \nonumber \\
&=&\left( E-H\left( \mathbf{r}\right) \right) ^{-1}\left( E-H\left( \mathbf{r%
}\right) +V\left( \mathbf{r}\right) \right) g(\mathbf{r,r}^{\prime }) 
\nonumber \\
&=&g(\mathbf{r,r}^{\prime })+\left( E-H\left( \mathbf{r}\right) \right)
^{-1}V\left( \mathbf{r}\right) g(\mathbf{r,r}^{\prime }).
\end{eqnarray}
From the inverse operator notation (\ref {Eq_op_inv_inte}) for $\left( E-H\left( \mathbf{r}\right) \right)
^{-1}$, $G(\mathbf{r,r}^{\prime })$ is
\begin{equation}
G(\mathbf{r,r}^{\prime })=g(\mathbf{r,r}^{\prime })+\int d\mathbf{r}_{1}G(%
\mathbf{r,r}_{1})V\left( \mathbf{r}_{1}\right) g(\mathbf{r}_{1}\mathbf{,r}%
^{\prime })  \label{Eq_Dyson_G}.
\end{equation}
This result is called the Dyson equation, and it allows us to express
the Green's function of the perturbed system in terms of the Green's function of the unperturbed system. Similarly, we can
find $ \psi \left (\mathbf{r}\right) $ from $ G (\mathbf{r, r}^{\prime}) $, for this we do
\begin{eqnarray}
E\psi _{0}(\mathbf{r})-H_{0}(\mathbf{r})\psi _{0}(\mathbf{r}) &=&0  \nonumber
\\
\left( E-H(\mathbf{r})\right) \psi _{0}(\mathbf{r}) &=&-V(\mathbf{r})\psi
_{0}(\mathbf{r}),
\label{Eq_Psi_0_V}
\end{eqnarray}%
in this case, the external source is given by $ f \left(\mathbf{r}\right) = -V \left(\mathbf{r}\right)\psi_{0}(\mathbf{r})$, and using (\ref{Eq_G_dif}), we can express
\begin{equation}
\psi _{0}(\mathbf{r})=\psi (\mathbf{r})-\int G(\mathbf{r},\mathbf{r}^{\prime
})V(\mathbf{r})\psi _{0}(\mathbf{r})d\mathbf{r}^{\prime },
\end{equation}%
where we use the solution of equation (\ref{Eq_Psi_0_V})
for $ V \left (\mathbf {r}\right) = 0 $ that is $\psi(\mathbf{r})$,  we found
\begin{equation}
\psi (\mathbf{r})=\psi _{0}(\mathbf{r})+\int G(\mathbf{r},\mathbf{r}^{\prime
})V(\mathbf{r}^{\prime })\psi _{0}(\mathbf{r}^{\prime })d\mathbf{r}^{\prime
},
\label{Eq_Sol_Psi_Dyson_G}
\end{equation}
which allows us to find the quantum state of the perturbed system from the unperturbed wave function and the Green's function $G$ that is found from the unperturbed Green's function $g$ solving the equation  (\ref{Eq_Dyson_G}).

\section{\label{sec:level1} Dirac delta potential in one dimension.}

\begin{figure}[t]
\centering\includegraphics[width=0.65\columnwidth]{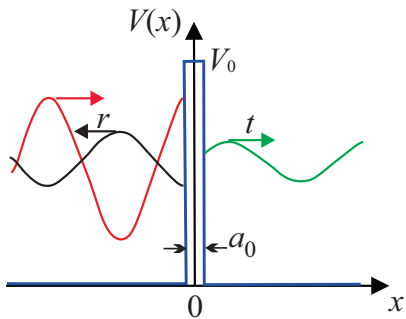}
\caption{Potential barrier of  width  $a_{0}$ and height $V_0$, when  $a_{0} \rightarrow 0$ and $V_{0} \rightarrow \infty$ but $a_{0}V_{0}\rightarrow U_{0}$ the potential can be modelled as $V\left( x\right) =U_{0}\delta \left( x\right)$. The red line illustrates the incoming wave, and the black and green lines are the reflected ($r$) and transmitted ($t$) waves, respectively.}
\label{fig:potencial}
\end{figure}

\begin{figure}[t]
\centering\includegraphics[width=0.7\columnwidth]{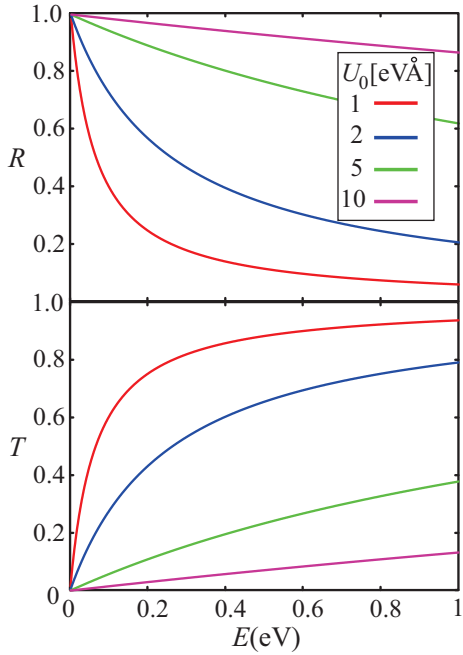}
\caption{Reflection and transmission coefficients as a function of the energy for different values of $U_0$. the units of $U_0$ are $eV \r{A}$ that corresponds to an insulating barrier for a typical metal-oxide-metal system.}
\label{fig:coeficientes}
\end{figure}

As a first example, we consider the Schr\"{o}dinger equation for a
particle in one dimension with a potential of the form
\begin{equation}
V\left( x\right) =U_{0}\delta \left( x\right), 
\end{equation}%
it can model a thin potential barrier that couples two
regions of a material, such as two metals separated by an oxide layer. Here, $ U_{0} = a_{0} V_{0}$, where $a_{0}$ is the width of the barrier and $V_0$, the height (see FIG. \ref{fig:potencial}), is a parameter that gives us the characteristic of how strong the barrier is. Dyson's equation in terms
of the coordinates $ x $ and $ x^{\prime} $, is written as

\begin{equation}
G(x,x^{\prime })=g(x,x^{\prime })+\int g(x,x_{1})V\left( x_{1}\right)
G(x_{1},x^{\prime })dx_{1},
\end{equation}
replacing the potential $ V\left(x\right) $, we get
\begin{equation}
G(x,x^{\prime })=g(x,x^{\prime })+U_{0}g(x,0)G(0,x^{\prime }).
\label{FG_1p_Dyson_deltaX}
\end{equation}
From this equation we can find $ G(0, x^{\prime})$
\begin{eqnarray}
G(0,x^{\prime }) &=&g(0,x^{\prime })+U_{0}g(0,0)G(0,x^{\prime })  \nonumber
\\
G(0,x^{\prime }) &=&\left[ 1-U_{0}g(0,0)\right] ^{-1}g(0,x^{\prime }),
\end{eqnarray}%
and replacing in (\ref{FG_1p_Dyson_deltaX}), we get
\begin{equation}
G(x,x^{\prime })=g(x,x^{\prime })+U_{0}g(x,0)\left[ 1-U_{0}g(0,0)\right]
^{-1}g(0,x^{\prime }).
\label{Eq_G_delta1}
\end{equation}
The  unperturbed function  $g(x, x^{\prime})$ corresponds to an infinite 
one-dimensional system. In Appendix A we  show  the  method  of  the  asymptotic  solutions  to calculate  the  Green’s  function  in  one  dimension  and, we applied  this  to  find  the  GF  of  the  free  particle, which is given by 
\begin{equation}
g(x,x^{\prime })=Ae^{ik\left\vert x-x^{\prime }\right\vert },
\label{FG_1p_FG_1D}
\end{equation}%
with 
\begin{equation}
k=\sqrt{2mE/{\hbar ^2}},  A = -i \frac{m}{\hbar^{2} k}.
\end{equation}
Replacing $g(x,x^{\prime })$ in Eq. (\ref{Eq_G_delta1}) we obtain,
\begin{equation}
G(x,x^{\prime })=A\left( e^{ik\left\vert x-x^{\prime }\right\vert
}+U_{0}Ae^{ik\left( \left\vert x\right\vert +\left\vert x^{\prime
}\right\vert \right) }\left[ 1-U_{0}A\right] ^{-1}\right) .
\end{equation}%
Defining $ Z $ as the strength of the barrier
\begin{equation}
Z=\dfrac{m}{\hbar ^{2}k}U_{0},
\end{equation}%   
we get,
\begin{equation}
G(x,x^{\prime })=A\left( e^{ik\left\vert x-x^{\prime }\right\vert }-\frac{iZ%
}{1+iZ}e^{ik\left( \left\vert x\right\vert +\left\vert x^{\prime
}\right\vert \right) }\right) .  \label{Eq_G_Z}
\end{equation}
The first term comes from the Green's function of the homogeneous system, and the second, from particle interaction with the potential, which breaks the homogeneity of the system.
 Now to
find the perturbed wave function, we use (\ref{Eq_Sol_Psi_Dyson_G}%
), with which
\begin{eqnarray}
\psi (x) &=&\psi _{0}\left( x\right) +\int G(x,x^{\prime })U_{0}\delta
(x^{\prime })\psi _{0}(x{^{\prime }})dx^{\prime }  \nonumber \\
&=&\psi _{0}\left( x\right) +G(x,0)U_{0}\psi _{0}(0).  \label{Eq_Psi_G_delta}
\end{eqnarray}
If we assume an incident wave from the left, the 
unperturbed wave function is
\begin{equation}
\psi _{0}\left( x\right) =ae^{ikx}  \label{Eq_psi_0},
\end{equation}%
with $ a $ the amplitude of the wave. From (\ref{Eq_G_Z}), we can express
\begin{equation}
G(x,0)=Ae^{ik\left\vert x\right\vert }\left( \frac{1}{1+iZ}\right) ,
\label{Eq_G_X0}
\end{equation}%
Using (\ref{Eq_Psi_G_delta}), (\ref{Eq_psi_0}) and (\ref{Eq_G_X0}), we obtain
\begin{equation}
\psi (x)=a\left( e^{ikx}-e^{ik\left\vert x\right\vert }\frac{iZ}{1+iZ}\right).
\end{equation}%
We can write explicitly to the left and right of the
barrier the wave function as
\begin{equation}
\psi (x)=\left\{ 
\begin{array}{cr}
a\left( e^{ikx}-e^{-ikx}\frac{iZ}{1+iZ}\right), & x<0   \\ 
ae^{ikx}\frac{1}{1+iZ} , & x>0. %
\end{array}%
\right. 
\end{equation}%
From here, we can see that the amplitudes of reflection ($ r $), and
transmission ($ t $) are
\begin{equation}
r=\frac{-iZ}{1+iZ}, \hspace{0.2cm} t=\frac{1}{1+iZ},
\end{equation}%
in such a way that the reflection and transmission coefficients \footnote {%
In general, the transmission coefficient is defined as the ratio
between the transmitted probability current density and the density of
incident current. When the system is three-dimensional, it must
consider only the current densities normal to the interfaces} are
\begin{equation}
R=\left\vert r\right\vert ^{2}=\frac{Z^{2}}{1+Z^{2}},\;T=\left\vert
t\right\vert ^{2}=\frac{1}{1+Z^{2}}.
\end{equation}%
These coefficients satisfy that $ R^{2} + T^{2} = 1 $. These are
illustrated in FIG. \ref{fig:coeficientes}.\\
Let us go back to Green's function and analyze the case of an infinite barrier, for which we do $ U_{0}, Z \rightarrow \infty $

\begin{equation}
G(x,x^{\prime })=A\left( e^{ik\left\vert x-x^{\prime }\right\vert
}-e^{ik\left( \left\vert x\right\vert +\left\vert x^{\prime }\right\vert
\right) }\right) .  \label{eQ_G_Semi}
\end{equation}%
This Green's function corresponds to that of a semi-finite medium to
left or right at $ x = 0, $ and it will be used in the next section.
\bigskip
\\
\begin{figure}[t]
\centering\includegraphics[width=1\columnwidth]{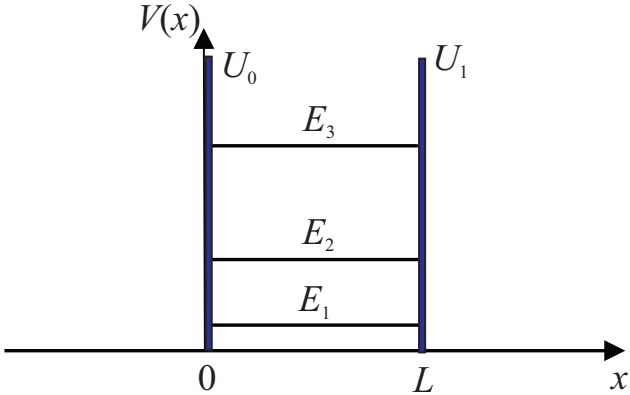}
\caption{Potential well, formed from two Dirac functions at
$x = 0$ and $x = L$, with  $U_0$ and $U_1$ as parameters. When $U_{0/1} \rightarrow 
\infty $, we obtain an infinite square potential well.}
\label{fig:potencialwell}
\end{figure}

\section{\label{sec:level3} Green's function for a quantum potential well.}

\begin{figure}[t]
\centering\includegraphics[width=1.05\columnwidth]{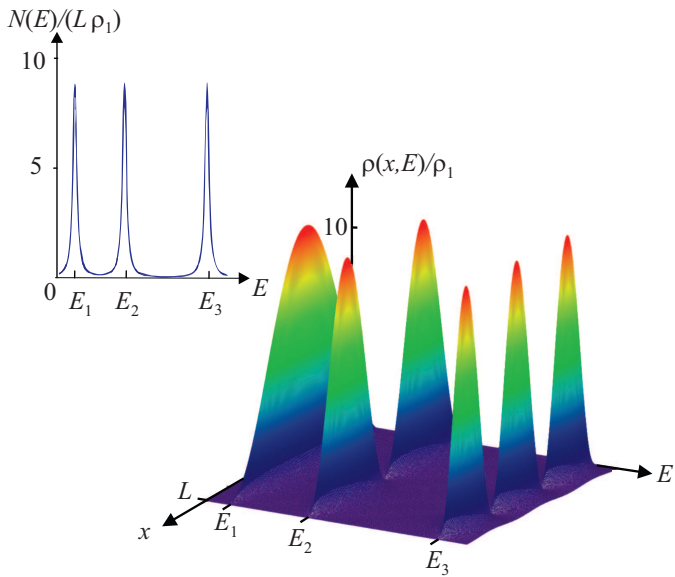}
\caption{Local density of states for an infinite square potential well, as a function of $x$ and energy. The inset shows the density of states, 
where we can observe that $N(E)$ is maximal when $E=E_{n}$. Here, $\rho_1=1/(2E_1L)$ and the integral of $N(E) dE$ for each peak is 1.}
\label{fig:LDOSwell}
\end{figure}

We are going to find the Green's function of an infinite potential well; see FIG. \ref{fig:potencialwell}.
To do this, we start from the Green's function of the semi-infinite medium with $ x> 0 $ and place an additional potential at $ x = L $
\begin{equation}
V\left( x\right) =U_{1}\delta (x-L),
\end{equation}%
with $U_{1}=a_{1}V_{1}$, $a_{1}$ being the width of the additional barrier and $V_1$ its height. For the Dyson equation the unperturbed Green's function is given by (\ref{eQ_G_Semi}) with $ x, x^{\prime}> 0 $ and denoted as $%
g (x, x^{\prime}) $
\begin{equation}
g(x,x^{\prime })=A\left( e^{ik\left\vert x-x^{\prime }\right\vert
}-e^{ik\left( x+x^{\prime }\right) }\right) .  \label{Eq_g_semi}
\end{equation}%
Proceeding similarly to the case of a barrier we have,
\begin{equation}
G(x,x^{\prime })=g(x,x^{\prime })+U_{1}g(x,L)\left[ 1-U_{1}g(L,L)\right]
^{-1}g(L,x^{\prime }),
\end{equation}
at the limit of $ U_ {1} \rightarrow \infty $, we obtain the Green's function 
 from an infinite square potential well as
\begin{equation}
G(x,x^{\prime })=g(x,x^{\prime })-g(x,L)\left[ g(L,L)\right]
^{-1}g(L,x^{\prime }).
\end{equation}
For $ x <x^{\prime} $ explicitly replacing (\ref{Eq_g_semi}), we get
\begin{equation}
G(x,x^{\prime })=-i\dfrac{m}{\hbar ^{2}k}\frac{\left(
e^{-ikx}-e^{ikx}\right) \left( e^{ikx^{\prime }}-e^{-ik(x^{\prime
}-2L)}\right)}{1-e^{i2kL}} .
\label{Eq_Well_pot}
\end{equation}
\\
From the poles of the Green's function, see Appendix $B$, we find that the energy spectrum is given by
\begin{equation}
e^{i2kL}=1,
\end{equation}
therefore 
\begin{equation}
\;k=\dfrac{n\pi }{L},\;E_{n}=\dfrac{\hbar ^{2}\pi ^{2}}{2mL^{2}}n^{2}.
\end{equation}
Which coincides with the eigenstates of an infinite square potential well. To
starting from $ G (x, x^{\prime}) $, we can find the local density of states,  see Appendix $B$, 
\begin{equation}
\rho \left( x,E\right) =-\frac{1}{\pi }\text{Im}\left[ G^{r}(x,x,E)%
\right],  
\end{equation}
remembering that  $ G ^ {r} (x, x, E) = G (x, x, E + i \eta) $ is the retarded Green's function. The figure \ref{fig:LDOSwell} illustrates the LDOS and  the density of states which is defined as
\begin{equation}
N \left(E\right)=\int dx \rho \left( x,E\right).
\end{equation}
Then, when $ E $ coincides with an eigenvalue of the system, $N(E)$ has a maxima, and $ \rho(x, E)$ is proportional to the probability density corresponding to that eigenvalue. 
\\ 

\section{\label{sec:leve5}Conclusions}
In this work, we have introduced the time-independent Green’s Function for the Schr\"{o}dinger equation. Also, we have derived the Dyson Equation to solve an ``inhomogeneous” Schr\"{o}dinger equation when an external potential perturbs the system. As a first example for applying the Dyson Equation with unbounded states, we solve a potential modelled by a Dirac delta function, where we find the Green’s function and the wave function, such as the reflection and transmission coefficients. To study bound states, we solve the infinite square potential well and find the GF, the energy spectrum, and local density of states. We showed the method of the asymptotic solutions to calculate the Green’s function in one dimension and applied this to find the GF of the free particle. The methods to find Green’s functions and examples show how to introduce this concept in quantum mechanics courses.

\begin{acknowledgments}
We wish to acknowledge the support of Universidad Nacional de Colombia, DIEB,  C\'odigos Hermes 48528 and 48148.
\end{acknowledgments}

\appendix

\section{Calculation of Green's function using 
asymptotic solutions}

\begin{figure}[t]
\centering\includegraphics[width=0.65\columnwidth]{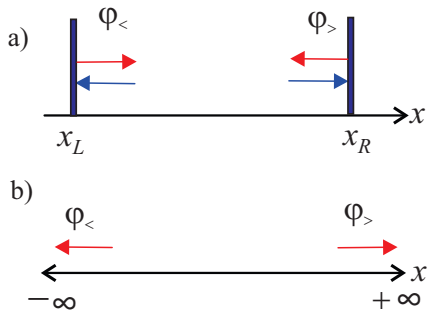}
\caption{Schematic representation of the asymptotic functions, the arrows illustrate the group velocity of each wave travelling in different directions. a) For a semi-infinite system with an edge at $x_{L}$ or $x_{R}$ and the function in each case is a linear combination of incoming (blue arrows) and outgoing (red arrows) plane waves. b) For an infinite system where $x_{R/L} \rightarrow \pm\infty$ and the functions are outgoing plane waves.}
\label{fig:asimp}
\end{figure}
One method to find the Green's function in one dimension is
to use the asymptotic solutions of the differential equation. As $
G(x, x^{\prime}) $ fulfills the homogeneous equation for $ x\neq x^{\prime
} $, we can write\begin{equation}
G(x,x^{\prime })=\left\{ 
\begin{array}{cl}
A\varphi_{<}(x)\varphi_{>}(x^{\prime }),& x<x^{\prime } \\ 
A^{\prime }\varphi_{<}(x^{\prime })\varphi_{>}(x),& x>x^{\prime }.%
\end{array}%
\right.   \label{ApexSchrMat_47_}
\end{equation}%
The functions $ \varphi_{<} ({x}) $ and $ \varphi_{>} ({x}) $ are solutions of the Schr\"{o}dinger equation that satisfy the boundary conditions
left or right respectively. For example, $\varphi_{<} (x_L) = 0 $, $ \varphi_{>} (x_R) = 0 $, $ x_L <x_R $, see FIG. \ref{fig:asimp}a. The product $\varphi_{<}(x)\varphi_{>}(x^{\prime })$ assures us that $ G (x, x^{\prime}) $ satisfies the Schr\"{o}dinger equation for both the operator $ H (x) $ as for $ H ({x}^{\prime}) $. The constants $ A $ and $ A^{\prime} $ are determined from the boundary conditions at  $x = x^{\prime} $. Integrating the differential equation
\begin{equation}
\left( E-V(x)+\dfrac{\hbar ^{2}}{2m}\dfrac{d^{2}}{dx^{2}}\right)
G(x,x^{\prime })=\delta (x-x^{\prime })  \label{ApexSchrMat_48_}
\end{equation}%
between $ x^{\prime} - \epsilon $ and $ x^{\prime} + \epsilon $, we get
\begin{equation}
\int\limits_{x`-\varepsilon }^{x`+\varepsilon }dx\left( E-V(x)+\dfrac{\hbar
^{2}}{2m}\dfrac{d^{2}}{dx^{2}}\right) G(x,x^{\prime
})=\int\limits_{x`-\varepsilon }^{x`+\varepsilon }dx\delta (x-x^{\prime }).
\label{ApexSchrMat_49_}
\end{equation}%
Taking the limit when $ \epsilon \rightarrow 0 $, we find that
\begin{equation}
\dfrac{dG(x,x+\varepsilon )}{dx}-\dfrac{dG(x,x-\varepsilon )}{dx}=\dfrac{2m}{%
\hbar^{2} },  \label{ApexSchrMat_50_}
\end{equation}%
integrating again
\begin{equation}
G(x,x+\varepsilon )=G(x,x-\varepsilon ).  \label{ApexSchrMat_51_}
\end{equation}%
Replacing the assumption \eqref{ApexSchrMat_47_} in the boundary conditions of \eqref {ApexSchrMat_50_} and \eqref {ApexSchrMat_51_}, we obtain
%\bigskip $%
\begin{eqnarray*}\centering
{A\varphi _{<}(x)\varphi _{>}(x)}{=A^{\prime }\varphi _{<}(x)\varphi _{>}(x)}
&& \\
{A^{\prime }\varphi _{<}(x)\dfrac{d\varphi _{>}(x)}{dx}-A^{\prime }\dfrac{%
d\varphi _{<}(x)}{dx}\varphi _{>}(x)}{=\dfrac{2m}{\hbar^{2} }} &&.
\end{eqnarray*}\\
Solving these equations 
\begin{equation}
A=A^{\prime } =\dfrac{2m}{\hbar^{2} W},%
\label{ApexSchrMat_52_}
\end{equation}%
where $W$ is the Wronskian,
\begin{equation}
W=\varphi _{<}(x)\dfrac{d\varphi _{>}(x)}{dx}-\dfrac{d\varphi _{<}(x)%
}{dx}\varphi _{>}(x),   \label{ApexSchrMat_53_}
\end{equation}%
which is independent of $ x $, $dW/dx=0$. With this, we finally write

\begin{equation}
G(x,x^{\prime })=\dfrac{2m}{\hbar^{2} W}\left\{ 
\begin{array}{l}
{\varphi_{<}(x)\varphi_{>}(x^{\prime }),\;x<x^{\prime }} \\ 
{\varphi_{<}(x^{\prime })\varphi_{>}(x),\;x>x^{\prime }.}%
\end{array}%
\right.  
\label{Eq_G_asymp}
\end{equation}%

As an example, we consider the free particle case which the asymptotic solutions are, see FIG. \ref{fig:asimp}b 
\begin{eqnarray*}
\varphi _{<}(x) &=&e^{-ikx}, \\
\varphi _{>}({x}) &=&e^{ik{x}},
\end{eqnarray*}%
with $ k = \sqrt{2mE / \hbar^2} $. The wave function $ \varphi_{<} (x) $ describes an electron that propagates to the left and $ \varphi_{>} (x) $ one that propagates to the right,
using (\ref{ApexSchrMat_53_})
\[
W=ike^{-ikx}e^{ik{x}}+ike^{-ikx}e^{ik{x}}=2ik,
\]%
and from Eq. (\ref{Eq_G_asymp}) we obtain
\begin{equation}
G(x,x^{\prime })=-i\dfrac{m}{\hbar^{2}k }\left\{ 
\begin{array}{ll}
e^{-ikx}e^{ik{x}^{\prime}},& x<x^{\prime}\\
e^{-ikx^{\prime }}e^{ik{x}},& x>x^{\prime},
\end{array}%
\right.  
\end{equation}%
which can be written as 
\begin{equation}
G(x,x^{\prime })=-i\dfrac{m}{\hbar^{2} k}e^{ik|x-x^{\prime }|}.
\end{equation}
\\

\section{Density of states}

For a system, such as a metal, in which the electronic energy levels form a quasi-continuum, a function of great importance is the density of states $ N(E) $, where $ N(E){dE} $ is defined as the number of states with energies between $ E $ and $ E + dE $. We are going to show that the density of states can be expressed in terms of a 
Green's function, then we are going to  find an expression of the GF in terms of the eigenfunctions of the Hamiltonian $ H $, which we assume orthonormal and complete
\begin{eqnarray}
\int d\mathbf{r}\varphi _{n}\left( \mathbf{r}\right) \varphi _{m}^{\ast
}\left( \mathbf{r}\right)  &=&\delta _{n,m},  \label{Eq_Ortog} \\
\;\sum\limits_{n}\varphi _{n}\left( \mathbf{r}\right) \varphi _{n}^{\ast
}\left( \mathbf{r}^{\prime }\right)  &=&\delta \left( \mathbf{r-r}^{\prime
}\right) .  \label{Eq_completez}
\end{eqnarray}%
Assuming a solution of the form
\begin{equation}
G\left( \mathbf{r},\mathbf{r}^{\prime }\right) =\sum\limits_{n}d_{n}\varphi
_{n}\left( \mathbf{r}\right) \varphi _{n}^{\ast }\left( \mathbf{r}^{\prime
}\right) ,
\end{equation}%
we replace it in
\begin{equation}
\left( E-H\right) G\left( \mathbf{r},\mathbf{r}^{\prime }\right) =\delta
\left( \mathbf{r-r}^{\prime }\right) ,
\end{equation}%
and using the completeness condition (\ref{Eq_completez}), we obtain
\begin{equation}
\sum\limits_{n}d_{n}\left( E-E_{n}\right) \varphi _{n}\left( \mathbf{r}%
\right) \varphi _{n}^{\ast }\left( \mathbf{r}^{\prime }\right)
=\sum\limits_{n}\varphi _{n}\left( \mathbf{r}\right) \varphi _{n}^{\ast
}\left( \mathbf{r}^{\prime }\right) 
\end{equation}%
in this way, $ d_{n} = \left(E-E_{n}\right)^{-1}, $ and therefore
\begin{equation}
G\left( \mathbf{r},\mathbf{r}^{\prime }\right) =\sum\limits_{n}\frac{%
\varphi _{n}\left( \mathbf{r}\right) \varphi _{n}^{\ast }\left( \mathbf{r}%
^{\prime }\right) }{E-E_{n}}.
\end{equation}%
To avoid singularity at $ E = E_{n} $ we define the retarded Green's function, as
\begin{equation}
G^{r}\left( \mathbf{r},\mathbf{r}^{\prime }\right) =\sum\limits_{n}\frac{%
\varphi _{n}\left( \mathbf{r}\right) \varphi _{n}^{\ast }\left( \mathbf{r}%
^{\prime }\right) }{E-E_{n}+i\eta },
\label{Eq_G_def_ret}
\end{equation}%
with $ \eta $ an infinitesimal part tending to zero, $ \eta \rightarrow 0. $ Here we see that $ G^{r}$ has the property that it diverges when $ E $ coincides with an eigenvalue of system energy $ E_{n} $. This property allows us to find the energy spectrum from the
poles of the Green's function. Taking $ \mathbf{r}^{\prime} = \mathbf{r}$ and making the integral in $ d\mathbf{r} $, we get
\begin{equation}
\int d\mathbf{r}G^{r,r}\left( \mathbf{r,}E\right) =\sum\limits_{n}\dfrac{1}{%
E-E_{n}+i\eta }.
\end{equation}
This sum is usually expressed by an integral with the definition of the density of states $ N(E)$
\begin{equation}
\int d\mathbf{r}G^{r}\left( \mathbf{r},\mathbf{r},E\right) =\int dE^{\prime }%
\dfrac{N(E^{\prime })}{E-E^{\prime }+i\eta },  \label{ApexSchrMat_30}
\end{equation}
with

\begin{equation}
N(E)=\sum\limits_{n}\delta \left( E-E_{n}\right) .
\label{Eq_Densidad_Estados_def}
\end{equation}%

The denominator in (\ref{ApexSchrMat_30}) \ can be written as

\begin{eqnarray*}
\frac{1}{E-E^{\prime }+i\eta } &=&\frac{E-E^{\prime }-i\eta }{\left(
E-E^{\prime }\right) ^{2}+\eta ^{2}} \\
&=&\frac{E-E^{\prime }}{\left( E-E^{\prime }\right) ^{2}+\eta ^{2}}-i\pi 
\frac{1}{\pi }\frac{\eta }{\left( E-E^{\prime }\right) ^{2}+\eta ^{2}}.
\end{eqnarray*}%
When $ \eta \rightarrow 0 $ the first term corresponds to the 
Cauchy principal part and the second to a Dirac delta function,
\begin{equation}
\frac{1}{E-E^{\prime }+i\eta }=P\frac{1}{E-E^{\prime }}-i\pi \delta \left(
E-E^{\prime }\right) .  
\label{Eq_Parte_principal}
\end{equation}%
With this, the equation (\ref{ApexSchrMat_30}) is
\begin{equation}
\int d\mathbf{r}G^{r}\left( \mathbf{r},\mathbf{r}\right) =P\int dE^{\prime }%
\dfrac{N(E^{\prime })}{E-E^{\prime }}-i\pi N\left( E\right) .
\end{equation}%
Since the principal part is real, we have to
\begin{equation}
N\left( E\right) =-\frac{1}{\pi }\text{Im}\int d\mathbf{r}G^{r}\left( 
\mathbf{r},\mathbf{r},E\right) .
\end{equation}%
We define the local density of states $ \rho\left (\mathbf{r}, E\right) $ from
\begin{equation}
N\left( E\right) =\int d\mathbf{r}\rho \left( \mathbf{r},E\right), 
\end{equation}%
with
\begin{equation}
\rho \left( \mathbf{r},E\right) =-\frac{1}{\pi }\text{Im}\left[ G^{r}\left( 
\mathbf{r},\mathbf{r},E\right) \right] .
\end{equation}%
Using the equations (\ref{Eq_G_def_ret}), (\ref{Eq_Densidad_Estados_def}), and (\ref{Eq_Parte_principal}) $ \rho \left (\mathbf{r}, E \right) $ can be expressed as
\begin{equation}
\rho \left( \mathbf{r},E\right) =\sum\limits_{n}\left\vert \varphi
_{n}\left( \mathbf{r}\right) \right\vert ^{2}\delta \left( E-E_{n}\right), 
\end{equation}
which shows that the local density of states is proportional to the
probability density.\\

\section{Challenge Problems} 
\subsection{Problem: Resonances in a double barrier potential. } 
Consider a double barrier potential, which is modelled by two Dirac delta barriers 
\begin{equation}
V\left( x\right) =U_{a}\delta (x)+U_{b}\delta (L).
\end{equation}
Show that by replacing this potential in Dyson's equation, we obtain
\begin{equation}
G(x,x^{\prime })=g(x,x^{\prime })+U_{a}G(x,0)g(0,x^{\prime
})+U_{b}G(x,L)g(L,x^{\prime }).
\end{equation}
From this equation, two equations can be obtained that connect the functions $ G(x, 0)$ and $ G(x, L) $. From them, find that
\begin{eqnarray}
G(x,x^{\prime }) &=&\frac{A}{1-r_{a}r_{b}e^{2ikL}}\{ \left(
1-r_{a}r_{b}e^{2ikL}\right) e^{ik|x-x^{\prime }|
}\nonumber \\
&+&r_{a}e^{ik|x|}\left( e^{ik|x^{\prime
}| }+r_{b}e^{ikL}e^{ik|L-x^{\prime }|}\right) 
\nonumber \\
&+&r_{b}e^{ik|x-L|}\left( e^{ik|L-x^{\prime
}|}+r_{a}e^{ikL}e^{ik|x^{\prime }|}\right)
\},  \label{Eq_FG_pozo_finito}
\end{eqnarray}
with $k=\sqrt{2mE/\hbar ^{2}}$, $r_{i}$ and $t_{i}$ the amplitudes of reflection and transmission for each barrier, $i=a,b$, given by
\begin{eqnarray}
r_{i} &=&\frac{-iZ_{i}}{ 1+iZ_{i} }=\left\vert r_{i}\right\vert
e^{i\alpha _{i}}, \\
t_{i} &=&\frac{1}{ 1+iZ_{i}}=\left\vert t_{i}\right\vert
e^{i\beta _{i}},
\end{eqnarray}%
where $Z_{i}$ is the strength for each barrier 
\begin{equation}
Z_{i}=\dfrac{m}{\hbar ^{2}k}U_{i}.
\end{equation}
From the Green's function calculate the wave function using the
equation (\ref{Eq_Sol_Psi_Dyson_G}), and assuming an incident wave $
\psi_0(x) = e^{ikx} $. For $ x>L $, get
\begin{equation}
\psi (x)=te^{ikx},
\end{equation}%
with $ t $ the transmission amplitude of the double barrier potential, given by
\begin{equation}
t=\frac{t_{a}t_{b}}{1-r_{a}r_{b}e^{2ikL}}.
\end{equation}
In the symmetric case $ Z_ {a} = Z_ {b}, $ show that
\begin{equation}
t=\frac{\left\vert t_{a}\right\vert ^{2}e^{2i\beta _{a}}}{1-\left\vert
r_{a}\right\vert ^{2}e^{2ikL+2i\alpha _{a}}},
\end{equation}%
when 
\begin{equation}
e^{2ikL+2i\alpha _{a}}=1,
\end{equation}
find that 
\begin{equation}
t=\frac{t_{a}^{2}}{1-\left\vert r_{a}\right\vert ^{2}}=e^{2i\beta _{a}},
\end{equation}
thus, the transmission coefficient is one
\begin{equation}
T=\left\vert e^{2i\beta _{a}}\right\vert ^{2}=1.
\end{equation}
That constitutes resonant tunnelling through quasi-bound states from the well. Plot the transmission coefficient for different values of $ Z $ and observe how the width of each resonance depends on $ Z $.\\

\subsection{Problem: Green's function of an infinite quantum potential well
by asymptotic solutions. }
Consider an infinite potential well with boundaries at $ x = $ 0, and $ x = L $. The asymptotic solutions, are
\begin{eqnarray*}
\varphi _{<}(x) &=&e^{-ikx}+ae^{ikx}, \\
\varphi _{>}({x}) &=&e^{ik{x}}+be^{-ik{x}},
\end{eqnarray*}%
with $ a $ and $ b $ the reflection amplitudes. From the 
boundary conditions of the wave functions $ \varphi_{<} $ at $ x = 0 $, and $ \varphi_{>} $ at $ x = L $ for \textquotedblleft fixed endpoints \textquotedblright \ $
\varphi_{<} (0) = \varphi_{>} (L) = 0 $, show that\\
\begin{equation}
a=-1,b=-e^{i2kL}.
\end{equation}
From $\varphi _{<}(x)$ and $\varphi _{>}(x)$ obtain that   
\begin{equation}
G(x,x^{\prime })=-i\dfrac{m}{\hbar ^{2}k}\frac{\left(
e^{-ikx}-e^{ikx}\right) \left( e^{ikx^{\prime }}-e^{-ik(x^{\prime
}-2L)}\right)}{(1-e^{i2kL})} ,
\end{equation}
for $x<x^{\prime }$.

It matches the one found using the Dyson equation, Eq. (\ref{Eq_Well_pot}).\\

\nocite{*}

\bibliography{GFbiblio}

\end{document}